\documentclass{wlscirep}

\usepackage[utf8]{inputenc}
\usepackage[T1]{fontenc}
\usepackage{mathptmx}
\usepackage{etoolbox}
\usepackage{graphicx}
\usepackage{xcolor}
\usepackage{tabularx}
\usepackage{xspace}
\usepackage{amsmath} 
\usepackage{placeins} 
\usepackage{hyperref}

\newcommand{\Jdc}{\ensuremath{J_\text{dc}}\xspace}
\newcommand{\Jc}{\ensuremath{J_\text{c1}}\xspace}
\newcommand{\Jcc}{\ensuremath{J_\text{c2}}\xspace}
\newcommand{\Jconf}{\ensuremath{J_\text{conf}}\xspace}
\newcommand{\fconf}{\ensuremath{f_\text{conf}}\xspace}
\newcommand{\J}[1]{#1~MA/cm$^2$}
\newcommand{\V}[1]{#1~m/s}

\newcommand{\Aex}{\ensuremath{A_\text{ex}}\xspace}

\newcommand{\smin}{\ensuremath{s_\text{min}}\xspace}
\newcommand{\smax}{\ensuremath{s_\text{max}}\xspace}
\newcommand{\sinf}{\ensuremath{s_\infty}\xspace}
\newcommand{\sinit}{\ensuremath{s_\text{i}}\xspace}

\newcommand{\titre}{Current-controlled periodic double-polarity reversals in a
spin-torque vortex oscillator}

\title{\titre}

\author[1]{Chloé Chopin}
\author[1]{Simon de Wergifosse}
\author[1]{Anatole Moureaux}
\author[1,*]{Flavio Abreu Araujo}

\affil[1]{Institute of Condensed Matter and Nanosciences, Universit\'{e} catholique de Louvain, Place Croix du Sud 1, 1348 Louvain-la-Neuve, Belgium}
\affil[*]{flavio.abreuaraujo@uclouvain.be}

\begin{abstract}
Micromagnetic simulations are used to study a spin-torque vortex oscillator excited by an out-of-plane dc current.
The vortex core gyration amplitude is confined between two orbits due to periodical vortex core polarity reversals.
The upper limit corresponds to the orbit where the vortex core reaches its critical velocity triggering the first polarity reversal which is immediately followed by a second one.
After this double polarity reversal, the vortex core is on a smaller orbit that defines the lower limit of the vortex core gyration amplitude.
This double reversal process is a periodic phenomenon and its frequency as well as the upper and lower limits of the vortex core gyration are controlled by the input current density while the vortex chirality determines the onset of this confinement regime.
In this non-linear regime, the vortex core never reaches a stable orbit and thus, it may be of interest for neuromorphic application, for example as a leaky integrate-and-fire neuron.
\end{abstract}

\begin{document}
\flushbottom
\maketitle

\section*{Introduction}
\begin{figure}
    \includegraphics{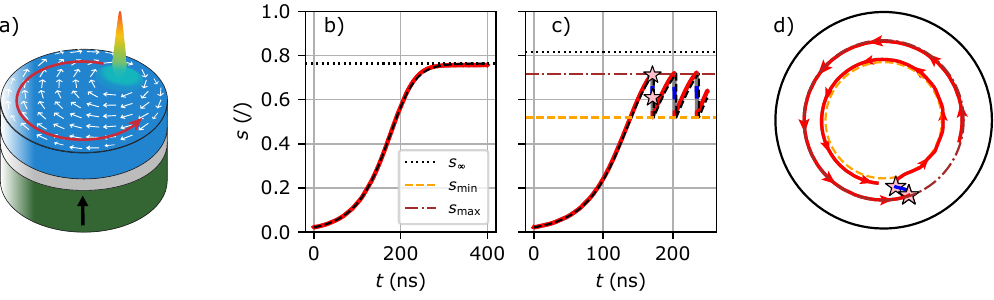}
    \caption{a) Schematic diagram of a STVO with the polarizer in green, insulator in gray and the free layer with a magnetic vortex as a ground state in blue. 
    b) Evolution of the reduced vortex core position $s(t)$ in the auto-oscillating regime saturating at $s_\infty$ with \Jdc$=$\J{-5.0}.  
    c) Evolution of $s(t)$ in the confinement regime with \Jdc$=$\J{-5.5}. 
    The trajectories of the vortex core given by micromagnetic simulations are plotted in red (resp.\ blue) for a positive (resp.\ negative) polarity.
    The black dashed lines are the estimation of $s(t)$ using  Eq.~(\ref{eq:s(t)}) or Eq.~(\ref{eq:conf}).
    Pink stars indicate polarity reversals. 
    d) Trajectory of the vortex core inside the magnetic dot while in the confinement regime.
    The vortex core motion direction is represented by arrows.
    When the polarity is negative, the gyration sense is reversed.}
    \label{fig:figure1}
\end{figure}

A magnetic vortex is a topological structure with a curling in-plane magnetization except at the vortex core where the magnetization points out-of-plane (see Fig.~\ref{fig:figure1}a).
Its polarity $P$ is positive ($P = +1$) when the out-of-plane (OOP) magnetization is pointing up and negative otherwise ($P = -1$).
Its chirality $C$ defines the curling in-plane magnetization orientation which is either clockwise ($C=-1$) or counterclockwise ($C=+1$).
These two parameters impact the vortex core gyrotropic motion as its polarity determines the sense of the vortex core gyration as well as the apparition of sustained oscillations while there is a splitting of its dynamics depending on its chirality\cite{choi2008understanding, araujo2021ampere}.
The vortex polarity can be reversed by  applying  different magnetic fields like static\cite{Okuno2002} or oscillating\cite{wang2012sub, yoo2012radial, ma2020periodic}  OOP magnetic fields as well as in-plane pulses\cite{xiao2006dynamics, Hertel2007}, oscillating\cite{van2006magnetic, lee2007ultrafast, choi2007strong, guslienko2008dynamic} or  rotating\cite{curcic2008polarization} magnetic fields.
The vortex polarity can also be reversed by a variety of spin-polarized currents like in-plane ac current\cite{yamada2007electrical, kim2007electric}  or current pulse\cite{liu2007current} as well as out-of plane dc current \cite{sheka2007current}.
Depending on the excitation, multiple polarity reversals can occur\cite{Hertel2007,yamada2007electrical,liu2007current} as in a nanocontact where a periodic polarity reversal is obtained with an input dc current~\cite{petit2012commensurability, Yoo_2020}.
The reversals occur in self-sustained oscillations leading to chaotic oscillations that may be interesting for secure communications\cite{petit2012commensurability}.

Here, a spin-torque vortex oscillator (STVO) is studied (see Fig.~\ref{fig:figure1}a).
The device is composed of two magnetic layers decoupled by a non-magnetic insulator. 
The first magnetic layer, the polarizer, has a fixed magnetization, while the second layer, the free layer, has a vortex as its ground state due to the magnetic dot's geometry\cite{Metlov_2002}. 
An input current density is injected in the STVO to trigger gyrotropic motion of the vortex core.
Indeed, the input current is spin-polarized by the polarizer and applies a spin-transfer torque on the free layer leading to sustained oscillations if the following conditions are met: $\Jdc P p_\text{z} < 0$ and $|\Jdc| \geq |\Jc|$ with \Jdc the input current density, $P$ the vortex polarization, $p_\text{z}$ the OOP magnetization of the polarizer and $\Jc$ the first critical current for sustained oscillations.
In a nanocontact, the Ampère-Oersted field (AOF) generated by the input current favours  a magnetic vortex, thus, the vortex chirality tends to align parallel to the AOF.
However, in a STVO, its initial chirality can be set deterministically\cite{Jenkins_2014}, thus its influence on the periodic reversal process can be studied. \\

Two distinct regimes\cite{guslienko2011spin, araujo2021ampere} arise depending on the vortex chirality and polarity, the polarizer orientation and the input current density: the resonant regime where the vortex core damps back to the magnetic dot center and the auto-oscillating regime where the vortex core reaches a stable orbit as in Fig.~\ref{fig:figure1}b.
Here, a third regime\cite{petit2012commensurability} is observed as shown in Fig.~\ref{fig:figure1}c and Fig.~\ref{fig:figure1}d.
Indeed, when the vortex core reaches the upper orbit \smax, its polarity is reversed.
This first polarity reversal is quickly followed by a second one, leaving the vortex core at the lower orbit \smin and in the auto-oscillating regime.
Since this double-reversals process is both periodical and sustained, the vortex core is confined between these two orbits.
This third regime is referred as the confinement regime and is only triggered for a vortex with a negative polarity.\\

\section*{Methods}
The vortex dynamics is studied by the means of micromagnetic simulations\cite{vansteenkiste2014design} using MuMax3.
The magnetic dot has a radius $R$ of 500~nm and a thickness $t$ of 9~nm.
The dot is discretized into cells of $2.5\times 2.5\times 4.5$~nm$^3$.
A magnetization saturation $M_s$ of 800~emu/cm$^3$ and an exchange stiffness\cite{guimaraes2017principles} \Aex of 1.07$\cdot10^{-6}$~erg/cm corresponding to permalloy are used.
The Gilbert damping constant $\alpha_\text{G}$ is set to 0.01 and the spin-current polarization is $p_J = 0.2$.
The polarizer is fixed along $+e_z$ while the initial vortex polarity is set to $P=+1$.
The temperature is set to $T =$~0~K and no external magnetic field is applied.
An input current density \Jdc between 0 and \J{-10} is applied to the STVO and the Ampère-Oersted field generated by the input current is added as an external magnetic field.

The evolution of the reduced vortex core position $s(t)$, with $s=||\mathbf{X}||/R$ and $\mathbf{X}$ the vortex core position, in the auto-oscillating regime extracted from MMS is fitted with the following equation\cite{moureaux2023neuromorphic}:
\begin{equation}
    s(t) = \dfrac{\sinit}{\sqrt{\left( 1 + \dfrac{\sinit^2}{\alpha/\beta}\right) e^{-2\alpha t} - \dfrac{\sinit^2}{(\alpha/\beta)}}}
    \label{eq:s(t)}
\end{equation}
with \sinit the initial position of the vortex core and two parameters $\alpha$ and $\beta$ that depend on \Jdc.
To predict $s(t)$ for a given \Jdc, 
the parameters $\alpha(\Jdc)$ and $\beta(\Jdc)$ are fitted with a linear fit: $\alpha(\Jdc) = a_\text{J} \Jdc + a$ and  $\beta(\Jdc) = b_\text{J} \Jdc + b$ when the vortex core is in the auto-oscillating regime, which gives $a_\text{J} =-$6.26~Hz~cm$^2$A$^{-1}$, $a=-12.75$ MHz, $b_\text{J}=5.42$~Hz~cm$^2$A$^{-1}$ and $b=-4.74$ MHz for a vortex with a negative chirality and $a_\text{J} =-12.49$~Hz~cm$^2$A$^{-1}$, $a=-13.47$ MHz, $b_\text{J}=$ 8.81~Hz~cm$^2$A$^{-1}$ and $b=3.00$ MHz for a vortex with a positive chirality.
In addition, the stable orbit \sinf is defined\cite{moureaux2023neuromorphic} by:
\begin{align}
        \sinf(\Jdc) =& \sqrt{-\dfrac{\alpha(\Jdc)}{\beta(\Jdc)}} 
        \label{eq:sinf}
\end{align}

To theoretically predict the evolution of $s(t)$ in the confinement regime the parameters \smin and \smax are needed as well as four hypotheses: 1) the vortex polarity is reversed when $s \geq \smax$; 2) it is always followed by a second polarity reversal; 3) the double reversal process is instantaneous and 4) after the second reversal $s = \smin$.
Both \smin and \smax can be fitted with a second order polynomial function.
The combination of Eq.~(\ref{eq:s(t)}), the four hypotheses and the expression of the parameters $\alpha, \beta, \smin \text{ and } \smax$ depending  on \Jdc  allow to predict $s(t)$ for any current density as in Fig.~\ref{fig:figure1}.
It gives the following equation with $\Delta t$ the time step, $s_n$ and $s_{n-1}$ the current and previous reduced vortex core position respectively:
\begin{equation}
        s_n = 
        \begin{cases}
            \dfrac{s_{n-1}}{\sqrt{\left( 1 + \dfrac{s_{n-1}^2}{\alpha/\beta}\right) e^{-2\alpha \Delta t} - \dfrac{s_{n-1}^2}{(\alpha/\beta)}}}, & \text{if } s_{n-1}<\smax \\
            \smin, & \text{if } s_{n-1} \geq \smax
        \end{cases}
    \label{eq:conf}
\end{equation}

\section*{Results and discussion}
The two configurations where the vortex chirality $C$ is either positive or negative are studied.
For  $|\Jc| \leq |\Jdc| \leq |\Jcc|$, with \Jc and \Jcc the first and second critical currents respectively,  the vortex core is in the steady-state regime. 
The evolution of the stable orbit $s_\infty$ with \Jdc depending on the vortex chirality is predicted thanks to Eq.~(\ref{eq:sinf}).
The first critical current density \Jc is defined by $\alpha(\Jc) = 0$ leading to $\Jc = -a/a_\text{J}$.
It gives $J_{\text{c1},C^+} =$\ \J{-1.08} and $J_{\text{c1},C^-} =$\ \J{-2.04} for a vortex with a positive and negative chirality respectively.

\begin{figure}[!ht]
    \centering
    \includegraphics[scale=1]{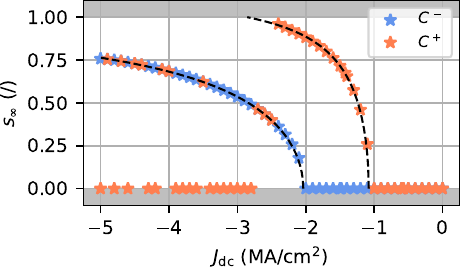}
    \caption{Evolution of the stable orbit $s_\infty$ with \Jdc for a positive or negative initial chirality. The black dashed lines are the prediction of $s_\infty(\Jdc)$ from Eq.~(\ref{eq:sinf}).}
    \label{fig:supp_chirality}
\end{figure}

The confinement regime is exclusively observed in vortices with negative chirality.
Indeed, for an initial positive chirality, the vortex core reaches the magnetic dot limit for $|\Jdc| \geq |\Jcc|$ with \Jcc$ \simeq $\J{-2.5}, its chirality is then reversed to $C=-1$ as it is favored by the AOF and  the vortex core polarity is either unchanged, leading to a steady-state regime similar to the one of a vortex with an initial negative chirality (see Fig.~\ref{fig:supp_chirality}), or reversed, resulting in a resonant regime.

\begin{figure*}[!ht]
    \centering
    \includegraphics{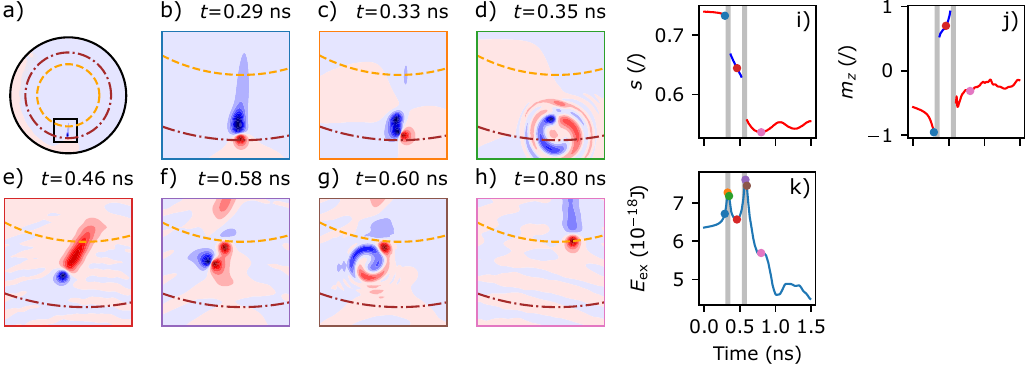}
    \caption{ a)-h) Evolution of OOP component of the vortex magnetization with time.
    a) View of the entire magnetic dot with the orbits \smin and \smax represented by orange dashed line and brown dashed-doted line respectively. 
    b)-h zoom  of the magnetization delimited by the black box in a).
    i) Evolution of $s$ as a function of time. The dot colours correspond to the border of the corresponding zoom from b) to h).
    Evolution of j) the dip magnitude $m_\text{z}$ and k) the exchange energy  $E_\text{ex}$ as a function of time.}
    \label{fig:Figure3} 
\end{figure*}

\begin{figure}
    \includegraphics{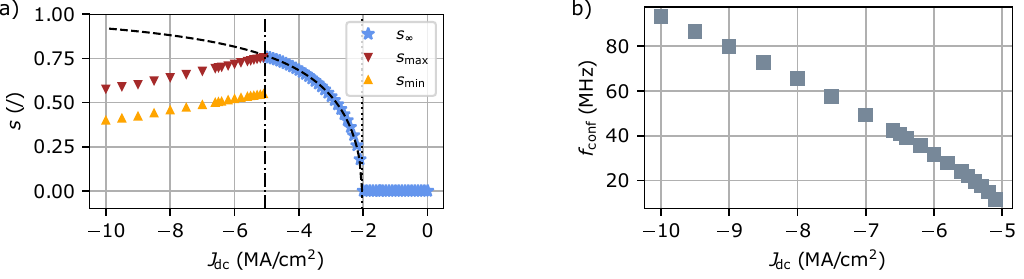}
    \caption{a) Evolution of $s_\infty$, $s_\text{min}$ and $s_\text{max}$ with $J_\text{dc}$.
    The dashed line is the prediction of  $s_\infty$ using Eq.~(\ref{eq:sinf}). The dotted and dashed-dotted lines are \Jc and \Jconf respectively.
    b) Evolution of $f_\text{conf}$ with $J_\text{dc}$.}
    \label{fig:figure2}
\end{figure}

For a vortex with a negative chirality, when $|\Jdc| \geq |\Jconf|$ with \Jconf$ \simeq $\J{-5.1}, the vortex core reaches its critical velocity at $s = \smax$ leading to its polarity reversal through the well-known process of creation and annihilation of a vortex-antivortex pair\cite{van2006magnetic, Hertel2007, xiao2006dynamics, sheka2007current, kim2007electric, lee2007ultrafast}.
Indeed, the vortex core reaches a velocity of \V{285} at \smax, which is coherent with the theoritical critical velocity\cite{Lee2008} $v_\text{cr} = \eta \gamma \sqrt{A_\text{ex}} = 302 \pm$ \V{32} with $\eta = 1.66 \pm 0.18$ and $\gamma = 2\pi\times2.8$~MHz/Oe, the gyromagnetic ratio.
As the velocity of the vortex core increases, it undergoes a deformation with the development of a dip with an opposite magnetization (see Fig.~\ref{fig:Figure3}).
At the critical velocity~\cite{Lee2008}, the dip amplitude is the same as the amplitude of the initial core polarization and the vortex polarity is reversed, as shown in Fig.~\ref{fig:Figure3}, through the creation and annihilation of a vortex-antivortex pair.
The excess of exchange energy is then dissipated through spin waves generation\cite{Hertel2006}.
As the polarity is then negative, the vortex core is in the resonant regime and relaxes towards its equilibrium position.
It experiences a second polarity reversal as the new dip reaches the same amplitude as the vortex core polarization and spin waves are once again generated. 
At the end of this second reversal, the vortex core is at $s_\text{min}$ and in the auto-oscillating regime again.
This double reversal happens periodically, leading to a sustained confinement regime described by three parameters: $s_\text{min}$ and $s_\text{max}$ that define the lower and upper limits of the reduced vortex core position, as well as the frequency $f_\text{conf}$ at which the double polarity reversals occur (see Fig.~\ref{fig:figure2}). 

The orbits \smin and \smax decrease as the amplitude of \Jdc increases as shown in Fig.~\ref{fig:figure2}a.
The time between two double-reversal is extracted and the corresponding confinement frequency \fconf is plotted in Fig.~\ref{fig:figure2}b.
The frequency increases with \Jdc, thus the double reversal frequency can be controlled by the dc input current with confinement frequencies starting at 11.5 MHz for \Jdc$=$\J{-5.1} to 93.5 MHz for \Jdc$=$\J{-10.0}.
Since spin waves are generated at each reversal, the frequency of spin wave generation is also controlled by \Jdc.

\FloatBarrier
\section*{Conclusion}
The dynamics of the vortex core is analytically described in both auto-oscillating and confinement regimes after fitting a few micromagnetic simulations.
The emergence of the confinement regime depends on the vortex chirality and \Jdc.
Therefore, by adjusting the latter, the parameters of the confinement regime, namely \smin, \smax, and \fconf, can be controlled, offering the potential to fine-tune the properties of the confinement regime as well as the generation of spin waves. 
Finally, the confinement regime exhibits nonlinear and periodic dynamics, making it potentially valuable for neuromorphic computing applications\cite{Grollier_2020} as the vortex core remains in a transient regime and never reaches a stable orbit.
It could be applied to leaky-integrate and fire neuron where the integration and leaking properties are controlled with the input current density and the firing correspond to the vortex core polarity reversal.
\FloatBarrier

\section*{Acknowledgements}
Computational resources have been provided by the Consortium des Équipements de Calcul Intensif (CÉCI), funded by the Fonds de la Recherche Scientifique de Belgique (F.R.S.-FNRS) under Grant No. 2.5020.11 and by the Walloon Region. F.A.A. is a Research Associate and S.d.W. is a FRIA grantee, both of the F.R.S.-FNRS.

\section*{Author's contribution}
C.C. and F.A.A. designed the study and analysed the results.
F.A.A. developped the analytical model.
The micromagnetic simulations were performed by C.C. and S.d.W..
C.C. wrote the core of the manuscript and all the other co-authors (S.d.W., A.M. and F.A.A.) contributed to the text as well as to the analysis of the results.

\end{document}